\UseRawInputEncoding 
\documentclass[pra,twocolumn,superscriptaddress]{revtex4}%
\usepackage{amssymb}
\usepackage{amsmath}
\usepackage{graphicx}
\usepackage{dcolumn}
\usepackage{xcolor}
\usepackage{bm}
\usepackage{subfigure}
\usepackage{amsfonts}
\usepackage{appendix}
\usepackage{tikz}

\usetikzlibrary{quantikz2,backgrounds,fit,decorations.pathreplacing}
\usepackage{xcolor}%
\providecommand{\U}[1]{\protect\rule{.1in}{.1in}}

\begin{document}
\preprint{APS/123-QED}

\title{Applications of Hybrid Machine Learning Methods to Large Datasets: A Case Study}

\author{Georgios Maragkopoulos}

\affiliation{Department of Informatics and Telecommunications, National and Kapodistrian
University of Athens, Panepistimiopolis, Ilisia, 15784, Greece}
\affiliation{Eulambia Advanced Technologies, Agiou Ioannou 24, Building Complex C, Ag. Paraskevi, 15342, Greece}
\author{ Nikolaos Stefanakos}

\affiliation{Department of Informatics and Telecommunications, National and Kapodistrian
University of Athens, Panepistimiopolis, Ilisia, 15784, Greece}
\affiliation{Eulambia Advanced Technologies, Agiou Ioannou 24, Building Complex C, Ag. Paraskevi, 15342, Greece}
\author{ Aikaterini Mandilara}

\affiliation{Department of Informatics and Telecommunications, National and Kapodistrian
University of Athens, Panepistimiopolis, Ilisia, 15784, Greece}
\affiliation{Eulambia Advanced Technologies, Agiou Ioannou 24, Building Complex C, Ag. Paraskevi, 15342, Greece}

\author{ Dimitris Syvridis}

\affiliation{Department of Informatics and Telecommunications, National and Kapodistrian
University of Athens, Panepistimiopolis, Ilisia, 15784, Greece}
\affiliation{Eulambia Advanced Technologies, Agiou Ioannou 24, Building Complex C, Ag. Paraskevi, 15342, Greece}

\begin{abstract}

We combine classical and quantum Machine Learning (ML) techniques to effectively analyze long time-series data acquired during experiments. Specifically, we demonstrate that replacing a deep classical neural network with a thoughtfully designed Variational Quantum Circuit (VQC) in an ML pipeline for multiclass classification of time-series data yields the same classification performance, while significantly reducing the number of trainable parameters. To achieve this, we use a VQC based on a single qudit, and encode the classical data into the VQC via a trainable hybrid autoencoder which has been recently proposed as embedding technique. Our results highlight the importance of tailored data pre-processing for the circuit and show the potential of qudit-based VQCs.

\end{abstract}
\maketitle

\section{Introduction}

Variational Quantum Circuits (VQCs)  refer to quantum circuits with parameterized gates, and they were first introduced as the core component of  variational quantum algorithms \cite{variational_quantum_algorithms}. Soon after, they were adopted in the field of Quantum Machine Learning (QML) \cite{machine_learning_with_quantum_computers, machine_learning_and_artificial_intelligence_in_quantum_domain} as the quantum analogue of classical Neural Networks (NNs) \cite{supervised_learning_with_quantum-enhanced_feature_spaces, quantum_machine_learning_in_feature_hilbert_spaces}. 
Today,  extensive numerical trials on benchmark Machine Learning (ML) problems and theoretical investigations suggest that VQCs are not yet sophisticated enough to outperform NNs. Moreover, the presence of barren plateaus \cite{barren_plateaus_in_qnn} poses significant challenges in training VQCs. On the other hand, research on the structure, capacity and training of  VQCs is still ongoing, and there are conjectures that VQCs may be better suited for handling quantum data or for solving specific problems \cite{Krems_2023}, much like quantum computers.
Additionally, the performance of VQCs has primarily been tested on feature spaces of small dimensions, often resulting from pre-processing of initial data with the PCA method.  Open questions thus remain on how VQCs perform on big unstructured data and also whether PCA as a pre-processing technique is the most suitable approach.

In this work, we employ a VQC for a multiclass classification task involving unstructured data acquired during experiments with a Quantum Key Distribution (QKD) system. In  previous work \cite{Maragkopoulos:2024QKD}, we performed the same task using a deep NN, providing a basis for comparison. Although the number of weights in the VQC is approximately one-sixtieth of those in the NN, their performance, as quantified by several metrics, is shown to be comparable.
To achieve this high level of performance with the VQC, we carefully design its circuit by replacing qubits with a qudit and by using pre-processing techniques \cite{Maragkopoulos:2024QAE} aimed at optimizing the embedding of classical data on the VQC. The results highlight the importance of these two aspects of a VQC: their adapted encoding within the VQC, and the power of qudits.

The manuscript is structured as follows. In Section~\ref{S2}, we review the acquisition and processing of data  to perform the classification task using purely classical methods. Specifically, we construct an ML pipeline, the final component of which is a deep NN, and we present the results of the classification achieved by this classical methodology on the test data. In Section~\ref{S3}, we review the main aspects of VQCs made of qudits and explain how they can function as Quantum Neural Networks (QNNs). The second key element of this work is the targeted encoding of classical data onto a VQC using a hybrid method, which we refer to as Quantum Autoencoders (QAEs). QAEs are described in Section~\ref{S4}. In Section~\ref{S5}, we replace the deep NN in the ML pipeline with a VQC composed of a nine-level qudit, where the inputs are first pre-processed by a QAE, and present the classification results achieved. This is the main model suggested in this work, and we refer to it as \textit{QAE--qudit VQC}.  For completeness, in the same section, we construct alternative hybrid models and compare their classification outcomes with those achieved by the QAE--qudit VQC model. In the Discussion section (Section~\ref{S6}), we summarize the key findings of this work and the elements of our suggested model.

\section{Data acquisition and a classical ML pipeline for their multiclass classification \label{S2} }

In a previous work \cite{Maragkopoulos:2024QKD}, we used a pair of QKD Toshiba terminals, QKD4.2A-MU and QKD4.2B-MU \cite{T12}, to collect Quantum Bit Error Rate (QBER) and Secure Key Rate (SKR) data \cite{used_data} under different conditions for the QKD fiber channel. These conditions include normal operation, coexisting classical signals at varying power levels, and different degrees of attenuation. More specifically, these different conditions define nine distinct labels for the data, which we briefly describe here as:
\begin{itemize}
    \item \textbf{Class 0:} Normal function
    \item \textbf{Class 1:} Coexistence using 1 Laser
    \item \textbf{Class 2:} Coexistence using 2 Lasers
    \item \textbf{Class 3-5:} Increased power levels of coexisting classical signals using 4 Lasers \& EDFA
    \item \textbf{Class 6-8:} Different degrees of attenuation on the quantum link.
\end{itemize}

The QBER/SKR data points are provided in sequential time steps, and thus, for each class, we have a long time-series of data. The goal was to develop a system that diagnoses the status of a QKD link in real-time, using only the latest $N$ QBER and SKR data provided by the system. To achieve this, we created $N$-plets of data from the long time-series. We used 80$\%$ of these $N$-plets, with $N=10$, for training, while the remaining $N$-plets were kept as test data.
In the last column of Table~\ref{tab:1}, we report the total number of $N$-plets acquired experimentally for each class. 
 
\begin{figure}[!t]
    \centering
 \includegraphics[width=0.9\linewidth]{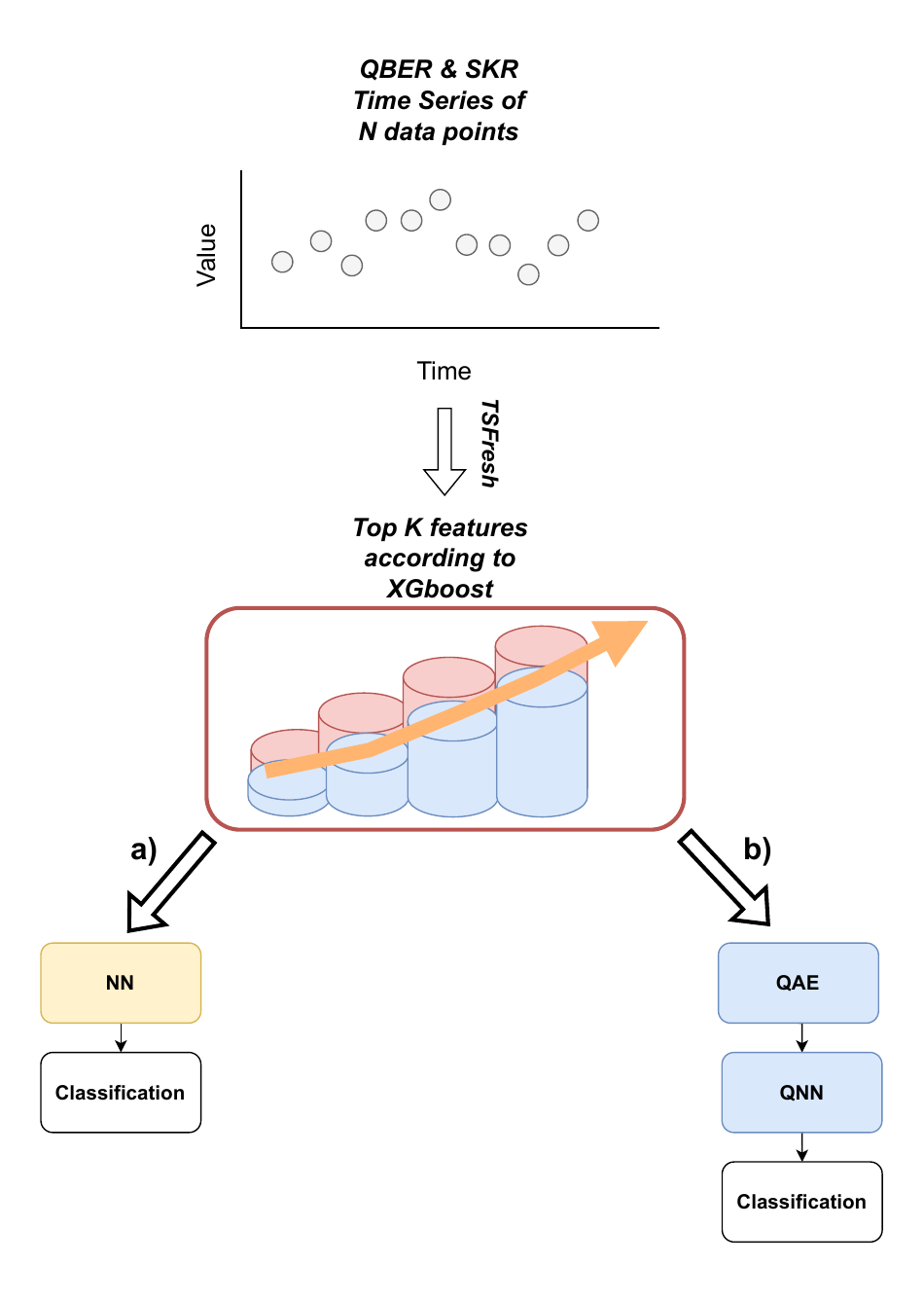}
    \caption{\textit{(a)} A schematic view of the ML pipeline suggested and tested in \cite{Maragkopoulos:2024QKD}. The aim is to classify
    the state of a QKD link, i.e., the impairments  on the quantum  channel,  by feeding
    to the ML pipeline the last $N$ QKD/SKR data. \textit{(b)} The modified hybrid ML pipeline that we test in the current work.  }
    \label{fig:1}
\end{figure}

To classify a test or new $N$-plet of data, a supervised learning model must first be trained. We built \cite{Maragkopoulos:2024QKD} such a model—an ML pipeline—by assembling \textit{tsfresh} \cite{MLQKD9}, \textit{XGBoost} \cite{MLQKD8}, and a deep NN model, as shown in Fig.~\ref{fig:1}~(a). The first component, \textit{tsfresh}, extracts various statistical, frequency-domain, and model-based features from the raw time series data ($N$-plets). From the $k$ extracted features, which are numerous ($k=1500$ for $N=10$), it is necessary to extract the most important for the classification task under study. For the selection of the  $K$ most important ones, we first train an XGBoost model using all the (labeled) train data. After this step,  each $N$-plet of QBER/SKR data is `mapped' to  $K$-plet with the values of these features estimated via \textit{tsfresh}.  In the final step, we use the (labeled) $K$-plets to train a deep NN with 3 hidden layers ($K \times 128 \times 256 \times 128 \times 9$) and ReLU activation functions. The output layer of the NN provides the probabilities for classifying an input data point, train, or test, into one of the nine classes.

We set $K=5$ and in Table~\ref{tab:1} we report the classification results from the ML pipeline ( Fig.~\ref{fig:1}~(a)) for the test QKD/SKR data. In the following sections, we gradually build a hybrid (quantum-classical) model by replacing the deep NN in the pipeline with a VQC, as illustrated in Fig.~\ref{fig:1}~(b).

\begin{table}[h!]
     \caption{ Metrics evaluating the classification results of the classical ML model schematically described in Fig.~\ref{fig:1}~(a). 
    The hyperparameters of the ML model  are chosen as $N=10$ and $K=5$.}\centering
    \label{tab:1}
    \footnotesize 
    \begin{tabular}{|c|c|c|c||c|}
        \hline
      Class $\#$ &   Precision & Recall & F1-Score &  $\#$ Data \\
        \hline
     \textbf{0}  & 0.89 & 1.00 & 0.94 &  4064 \\
        \hline
     \textbf{1}  &   1.00 & 0.10 & 0.18 & 395   \\
        \hline
     \textbf{2}  &   0.71 & 0.62 & 0.66 &  466 \\
        \hline
    \textbf{3}  &    0.98 & 0.95 & 0.96 &  392 \\
        \hline
     \textbf{4}  &    0.94 & 0.97 & 0.96 & 2070 \\
        \hline
     \textbf{5}  &    1.00 & 0.89 & 0.94 & 353  \\
        \hline
    \textbf{6}  &   0.97 & 1.00 & 0.98 & 1067  \\
        \hline
     \textbf{7} &   1.00 & 0.97 & 0.99 &  1529  \\
        \hline
     \textbf{8}  &    1.00 & 1.00 & 1.00 & 1629  \\
        \hline \hline
       & & & & \\
        \hline
       Accuracy &     0.94 &  & &  \\
        \hline
        Macro Average & 0.94 & 0.83 & 0.85 & \\
        \hline
      \end{tabular}
\end{table}

\section{VQCs made of qudits \label{S3}}

In circuit-based quantum computation, qubits serve as computational units, quantum gates implement the processing, and projective measurements are used to extract the outcomes. Due to considerations related to quantum error correction, the allowed operations are typically restricted to a universal set of gates, such as the set formed by the Hadamard, Phase, and CNOT gates.
VQCs are  quantum circuits which also include tunable quantum gates. The parameterized gates within the circuit allow it to explore a range of quantum computational models. This class of models is defined by the ansatz of the circuit, which refers to the choice of parameterized gates, fixed gates, and the architecture of the circuit.

A qubit describes the state of a discrete degree of freedom in a quantum system, which is quantized and yields two distinct outcomes when measured. Qudits, on the other hand, refer to quantum systems that provide $d$ distinct outcomes upon measurement and naturally generalize the concept of qubits. In this work, for reasons explained later in this section, we choose to use a VQC built from a qudit, rather than using multiple qubits. The following paragraphs provide the basic mathematical framework for these physical entities and their associated gates. Since a qudit with $d=2$ is equivalent to a qubit, this allows for an easy transition  to the well-known qubit case.

As a mathematical entity a qudit state `lives' in the $d$-dimensional Hilbert space which is spanned by the eigenstates of the Hamiltonian of the system. Let us denote by $\left\{\left|k\right\rangle\right\}_{k=0}^{d-1}$ such a set of normalized eigenstates and we refer to this set as the \textit{computational basis} of the qudit. Then one can
express a generic qudit state:
\begin{equation}
\left|\psi\right\rangle=\sum_{k=0}^{d-1} c_k \left|k\right\rangle ,
\end{equation}
by the complex amplitudes $c_i$'s over the computational basis, being constrained by the normalization condition $ \sum_{k=0}^{d-1} \left|c_k\right|^2 =1$.

Let us consider the full  $su(d)$ algebra for  the system, spanned by $D=d^2-1$ generators $\left\{\hat{g}_i\right\}$. For $d=2$ these generators can be identified with the Pauli operators while  for $d=3$  with the Gell-Mann operators. Employing the generators, we can conveniently express the most general unitary operator  $\hat{U}$ applied to a qudit  as:
\begin{equation}
 \hat{U}(\vec{\phi})=e^{-i \sum_{j=1}^{D}\phi_j\hat{g}_j},
 \label{eq:1}
\end{equation}
where we have ignored the global phase of $\hat{U}$ and $\phi_j$  are angles forming the vector $\vec{\phi}=\left\{\phi_1,\ldots,\phi_{D}\right\}$.
Eq. (\ref{eq:1}) also permits us to straightforwardly define a multi-parametrized gate with a set of parameters $\vec{\phi}$. One step further, we can employ (\ref{eq:1})  to construct  parametrized gates with a single parameter, which are commonly used in VQCs. If we define $\phi=\sqrt{\sum_j \phi_j^2 }$ as well as a normalized real vector $\vec{n}=\vec{\phi}/\phi$, then we can generate a parametrized gate as:
\begin{equation}
   \hat{R}_{\vec{n}}(\phi)= e^{-i \phi \vec{n}.\vec{\hat{g}}},
   \label{eq:2}
\end{equation}
where $\vec{\hat{g}}=\left\{\hat{g}_1,\ldots,\hat{g}_{D}\right\}$. Two-qudit (parametrized) gates  can be build in a similar way to single qudit gates but now one needs to replace the generators $\hat{g}_i$ in (\ref{eq:1}) by tensor products of generators, i.e., $\hat{g}_i \otimes \hat{g}_j$,  of the two qudit subsystems. 

As for qubits, measurable quantities on a qudit are described by operators represented by $d \times d$
Hermitian matrices. In this work for simplicity we employ observables with non-degenerate  spectrum.

With the introduced elements, one can construct a VQC made of qudits, similar to qubits, by using multiple qudits as registers, along with single- and two-qudit gates, as well as parameterized gates. In this work, we focus on using VQCs as QNNs, and in the next subsection, we explain the additional components necessary for a VQC to serve this purpose.

\subsection{VQCs as QNNs}
VQCs are one of the main models of QML, performing the functions of classical NNs in the quantum domain. Once the ansatz, i.e., gates and architecture, is chosen for a VQC, there are four further key points to address for it to function as a QNN: embedding classical features, performing read-out via measurements, constructing a loss function, and applying classical optimization methods for training.

There are four main methods \cite{machine_learning_with_quantum_computers} for performing the embedding of classical features $\vec{x}=\left\{x_1,\ldots,x_m\right\}$ in a VQC:  basis, amplitude, Hamiltonian and angle encoding. While amplitude encoding is the most efficient in terms of qubit resources,  angle encoding is the most widely used. In this work we  use \textit{parametrized} angle encoding  as:
\begin{equation}
 \hat{U}=e^{-i \sum_{j=1}^{m}x_j\phi_j\hat{g}_j}~.
 \label{eq:3}
\end{equation}
By inserting  trainable parameters $\vec{\phi}=\left\{\phi_1,\ldots,\phi_m\right\}$ in the embedding part of the circuit, one  can explore a bigger class of quantum feature maps: $\vec{x}\rightarrow \ket{\psi(\vec{x},\vec{\phi'})}$ where $\ket{\psi(\vec{x},\vec{\phi'})}$
the state-output of the quantum circuit. The vector $\vec{\phi'}$, as compared to  $\vec{\phi}$, includes also the  additional parameters of the variational part of the VQC.
Finally, we exercise the regular practice of re-uploading \cite{data_re-uploading_for_a_universal_quantum_classifier,supervised_learning_with_quantum-enhanced_feature_spaces} the classical features more than once in the VQC. Re-uploading
in a VQC increases the non-linearity of the quantum feature map, imitating the effect of hidden layers in NNs.
 
As for quantum computation,  VQC's output state  $\ket{\psi(\vec{x},\vec{\phi'})}$,   must be ``collapsed''
via measurement in order to derive classical information from it. To do this, one must first select the observable(s) $\hat{O}$ to be measured. 
The procedure — evolving via the VQC and performing measurement(s) — needs to be repeated multiple times  to estimate the expectation value(s),
$ \bra{\psi(\vec{x},\vec{\phi'})}\hat{O} \ket{\psi(\vec{x},\vec{\phi'})}$. This expectation value is the main input to the chosen loss function which is then optimized using a classical algorithm. In the common case where a classical gradient descent
method is employed for loss minimization, additional measurements are required to 
estimate  the gradient \cite{evaluating_analytic_gradients_on_quantum_hardware}.

In this work, we employ a VQC consisting of a single qudit. The potential advantages of replacing qubits with qudits in quantum technologies have been explored in several theoretical studies (see for instance \cite{Erhard2018, Mandilara2024, data_re-uploading_with_a_single_qudit}), while advancements in qudit technology, particularly in the field of photonics \cite{PhysRevLett.123.070505, Kues2017, Imany2019, qudit_Brien}, have been accelerated. Our main reason for using a qudit instead of multiple qubits is that it simplifies the circuit design, eliminating the need to optimize the placement of entangling and single-qubit gates. In a qudit, all operations, as described in (\ref{eq:1}), can be considered local and treated on an equal footing. Similarly, there is no need to combine measurement outcomes from multiple qubits; instead, we can directly work with single qudit's observables. Overall, qudits allow us to eliminate the need for optimization over the hyperparameters of the VQC and to explore the Hilbert space in a uniform way. The latter strategy seems the best option for the dataset under study, being unstructured.

\section{QAEs for embedding the classical features on VQCs \label{S4}}

In this work, we rely on the QAE model to pre-process classical data in a way that is compatible with VQCs. Originally introduced in earlier research \cite{Maragkopoulos:2024QAE}, a QAE incorporates the quantum feature map of the VQC under study into the bottleneck layer of a classical autoencoder. This design bridges classical and quantum techniques, providing a compact and meaningful representation of the input data.

The QAE operates similarly to a traditional autoencoder. Given an input dataset \(\mathbf{X} \in \mathbb{R}^{M \times K}\), where \(M\) is the number of samples and \(K\) is the number of features, the encoder compresses the data into a smaller latent space. Mathematically, this can be expressed as:  
\begin{equation}
\mathbf{Z} = f_\text{enc}(\mathbf{X}; \theta_\text{enc}),
 \label{eq:4}
\end{equation}
where \(\mathbf{Z} \in \mathbb{R}^{M \times D}\),  and \(f_\text{enc}\) represents the encoder function parameterized by weights \(\theta_\text{enc}\). In the QAE, the latent features \(\mathbf{Z}\) are then passed through a quantum feature map, which is implemented as a unitary transformation:
\begin{equation}
\hat{U} = \exp\left[-i \sum_{j=1}^{D} z_j \hat{g}_j \right].
 \label{eq:5}
\end{equation}
The quantum feature map is tightly related to the VQC's, reproducing its encoding layer. Technically, the feature map embeds the latent representation into the Hilbert space, encoding the data into unitary matrices. The  unitary operator in (\ref{eq:5}) is then applied to the ground state of the VQC circuit, $\ket{0}$. The real and imaginary values of the complex amplitudes of the resulting  vector, are passed onto the decoder, which reconstructs the original data, as $\mathbf{\tilde{X}}$, using classical layers.  The model is trained to minimize the reconstruction error, typically measured using the mean squared error (MSE):
\begin{equation}
\mathcal{L} = \frac{1}{M} \sum_{i=1}^M \|\mathbf{X}_i - \mathbf{\tilde{X}}_i\|^2.
 \label{eq:6}
\end{equation}
By minimizing \(\mathcal{L}\), the QAE ensures that the compressed ($D<K$) or uncompressed ( $D>K$) representation retains the most relevant features of the input data.

\section{Classification results via hybrid classical-quantum ML pipelines \label{S5}}

After presenting the basic elements of the VQC and QAE models, we proceed to employ them for the multi-class classification task described in Section \ref{S2}. We keep the first part of the ML pipeline intact and feed the K-plets of features into a QAE for pre-processing and then  to a single qudit VQC (see Fig. \ref{fig:1}~(b)). This hybrid model yields results comparable to those of the deep NN (Table \ref{tab:1}). To ensure the integrity of the model, we also construct two secondary models and present their classification outcomes. Specifically, the additional models are: \textit{a)} a qudit VQC without QAE, and \textit{b)} a VQC made of qubits assisted by QAE.

\subsection{QAE--qudit VQC model }

Let us consider  a qudit in the  state $\ket{\psi}=\sum_{k=0}^{d-1}c_k\ket{k}$  and describe the  measurement process of    an observable $\hat{O}$ with eigenstates the vectors of the computational basis  and  $d$ eigenvalues all different to each other. According to the axioms of quantum mechanics,  the measurement outcome related with the eigenvector
$\ket{k}$ occurs with probability $|c_k|^2$. We aim to solve a classification problem with $9$ labels
and it is natural to consider a qudit with $d=9$ and associate each  measurement outcome and its related probability with one of the nine classes. The data being unstructured guide us to employ all $D=80$ generators of the $SU(9)$ symmetry group and explore the Hilbert space of the qudit in a uniform way. We describe the \textit{basic  VQC unit}  by the unitary operation:
\begin{equation}
    \hat{U}(\vec{\phi}_l,\vec{x})=\mathrm{exp}\left[-i\sum_{j=1}^{D}x_j \phi_l^j \hat{g}_j\right],
    \label{eq:7}
\end{equation}
where $x_j$ the classical features and $\phi_l^j$ the trainable weights of this VQC unit. This way we choose the
simplest and most compact form for the VQC putting together the variational and encoding parts.
In order to increase the expressivity of the circuit we repeat the basic VQC unit, (\ref{eq:7}), $8$ times, re-uploading this way the features with different weights. The  total VQC contains  a total number of $640$ $\phi$ parameters and it is described by the unitary operator:
\begin{equation}
    \hat{U}_{VQC}=\hat{U}(\vec{\phi}_8,\vec{x})\ldots \hat{U}(\vec{\phi}_1,\vec{x}).
    \label{eq:8}
\end{equation}
As initial state for the qudit we consider its  ground state $\ket{0}$.

The final step involves training the VQC to identify the optimal weights  $\phi$ in the circuit (\ref{eq:8})  achieving the best classification outcomes for the train data. However, one may observe that the circuit is designed to accommodate $D=80$ features  $x_j$, while each data point is assigned $K=5$ features via tsfresh-XGBoost models. We address this discrepancy and further enhance the classification results by incorporating a pre-processing of $K$-plets of features via a QAE.

\begin{figure*}
    \centering
 \includegraphics[width=0.9\linewidth]{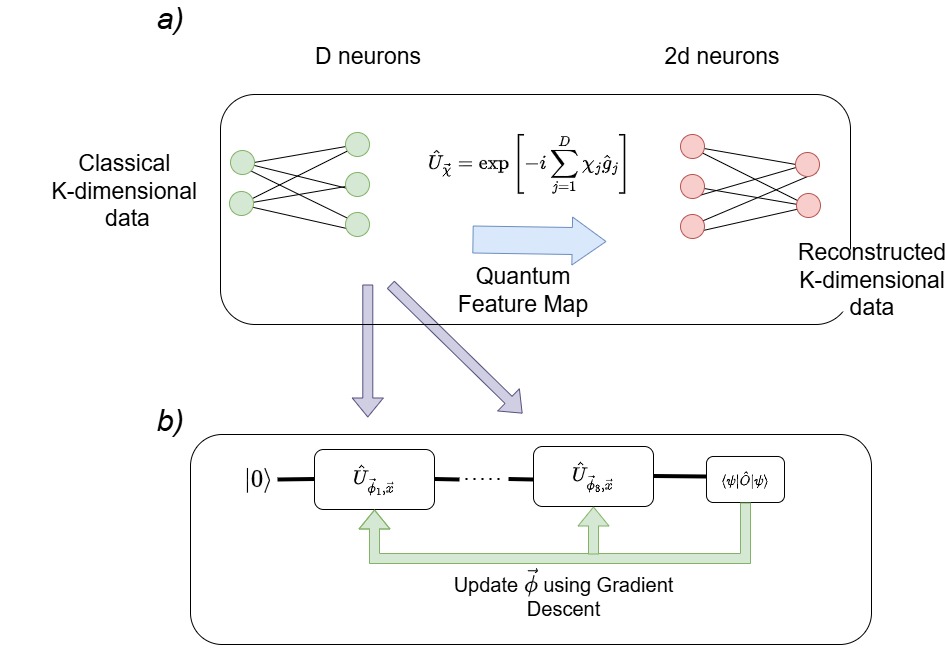}
    \caption{QAE--qudit VQC model. (a) A schematic representation of the QAE,  trained to create an adapted mapping: feature $K$-plets  to $D$-plets. For our case study, $K=5$ and $D=80$. This model consists of classical trainable encoding and decoding layers, with a quantum feature map in the bottleneck.
    The QAE is trained to minimize the MSE between input and output. After training, the output values $\chi$  from the D neurons (second layer) are used as input features to the VQC shown in (b). The VQC operates on a qudit with $d=9$ and is formed by a sequence of  basic VQC units, $\hat{U}(\vec{\phi}_l,\vec{x})$, with $l=1,\ldots 8$, (\ref{eq:7}). The $D$-plet of features provided by the QAE, is  re-uploaded on each of these  units. }
    \label{fig:2}
\end{figure*}
\textit{Pre-processing of data via QAE:} The aim of the QAE is to produce $D=80$ (redundant) features $x_j$ needed for the VQC (\ref{eq:8}), for each $K$-plet of features, with $K=5$.
The structure of the QAE that we use is presented in the Fig.~\ref{fig:2}~(a). The classical encoding part consists
by an input layer of $K=5$ neurons and a second layer of $D=80$ neurons. The neurons of the two layers   are fully connected with linear weights $\theta$. The $\chi_i$ outputs  of the second layer are used to encode the quantum layer of QAE and generate the state
\begin{equation}
    \ket{\chi}=\mathrm{exp}\left[-i\sum_{j=1}^{D} \chi_j \hat{g}_j\right]\ket{0}~~.
    \label{eq:9}
\end{equation}
We then use the $d=9$ complex amplitudes of the state $\ket{\chi}$ over qudit's computational basis to
create an input to the classical decoder consisting of a first layer of $2d= 18$ neurons and an output layer of $K$ neurons.
The two decoding layers are also fully connected to each other in a linear way. The QAE is trained with the train  data so that
MSE (\ref{eq:6}) is minimized. After this step, we use the QAE with the optimized weights $\theta$ to create for each $K$-plet of features, that is the input to the QAE, a sequence of $D$ features $x_j$ which are identified the outputs $\chi$ of the second classical layer.

The classification results for the test data, obtained using the combined method of QAE and qudit VQC, are reported in Table~\ref{tab:2}. These results are comparable to those achieved by the deep NN, as shown in Table~\ref{tab:1}. The number of trainable weights in the NN is approximately $ 6.7 \times 10^4$, while in the proposed QAE--qudit VQC model model, it is approximately $ 10^3$. Additionally, our proposed method decomposes into two independent optimization procedures—one for the QAE and one for the qudit VQC.  For these results, we numerically simulated the qudit VQC and successfully processed the considerably large dataset for QML standards \cite{SINGH2025114938,Haug_2023}, consisting of around $10,000$ rows, with a runtime approximately $\approx 10$ times slower than that of the deep NN.

\begin{table}[t]
    \caption{ Metrics evaluating the  classification results of the QAE--qudit VQC model. }
    \centering
    \label{tab:2}
    \footnotesize 
    \begin{tabular}{|c|c|c|c|}
        \hline
      Class $\#$ &   Precision & Recall & F1-Score  \\
        \hline
     \textbf{0}  & 0.91 & 0.96 & 0.93  \\
        \hline
     \textbf{1}  &   0.82 & 0.14 & 0.23    \\
        \hline
     \textbf{2}  &   0.87 & 0.47 & 0.61  \\
        \hline
    \textbf{3}  &    0.99 & 0.95 & 0.97  \\
        \hline
     \textbf{4}  &    0.93 & 1.00 & 0.96  \\
        \hline
     \textbf{5}  &    1.00 & 0.91 & 0.95  \\
        \hline
    \textbf{6}  &   0.96 & 1.00 & 0.98   \\
        \hline
     \textbf{7} &   0.88 & 0.99 & 0.93   \\
        \hline
     \textbf{8}  &    0.99 & 1.00 & 0.99   \\
        \hline \hline
       & & &  \\
        \hline
       Accuracy &     0.93 &  &   \\
        \hline
        Macro Average & 0.93 & 0.82 & 0.84  \\
        \hline
      \end{tabular}
\end{table}

\subsection{Secondary models of lower performance}

The results in Table \ref{tab:2} are encouraging, demonstrating very good performance for the proposed model on the given dataset. However, one might question whether the QAE and qudit structure of the VQC, as described in (\ref{eq:8}), are necessary components for these successful outcomes. It is straightforward to justify the use of QAE by comparing it to the results achieved without its inclusion. Therefore we implement the classification task with the qudit VQC of (\ref{eq:9}) but now we embed the $K$-plet of features, $K=5$, on the qudit VQC without pre-processing. The basic VQC unit is similar to (\ref{eq:8}) but  $75$ out of $80$ features need to be reset to unity: 
\begin{equation}
    \hat{U}(\vec{\phi}_l,\vec{x})=\mathrm{exp}\left[-i\sum_{j=1}^{5} x_j \phi_l^j \hat{g}_j-i\sum_{j=6}^{80} \phi_l^j \hat{g}_j\right].
    \label{eq:10}
\end{equation}
To eliminate the scenario where the choice of generators, $\hat{g}_i$, encoding the non-unity features is crucial, we randomly permute the ordering of the $\hat{g}_i$'s before applying equation (\ref{eq:10}). We perform the classification for three different random permutations, resulting in different VQC basic units, as shown in equation (\ref{eq:10}). The outcomes of these three runs are very similar, and in Table \ref{tab:3}, we report one of them.

Regarding the use of a qudit instead of qubits, this is a more complex question, as we would need to compare the results with the best outcomes achieved using a qubit-based VQC. To match the dimensions of the qudit's Hilbert space with a qubit circuit we  consider a VQC made of $4$ qubits. However, given the extensive design possibilities for a four-qubit circuit, this comparison becomes infeasible. Therefore, we provide an indicative result from a circuit that we designed as closely as possible to the qudit circuit.

Let us describe  the  qubit model consisting by a VQC that is assisted by a QAE. Each of the four qubits receives three features $x_i$  via the parametrized angle encoding transformation: $exp \left[ -i \sum_{j=1}^3 x_i \phi_i \hat{g}_i\right]$. The encoding is succeeded by  a cascade arrangement of three CNOT  gates which aims to entangle the qubits thus allowing the circuit to  explore the full dimension of the Hilbert space. Since we aim to employ re-uploading technique,  this basic VQC unit is repeated $m$ times. To match the given $K$-plet to the $12$ features of the qubit-VQC the circuit, and also to enhance the outcomes, we employ the `batched' QAE presented in Fig.~5~(b) of the work \cite{Maragkopoulos:2024QAE}. 
We have performed the classification task using $m=6,~8$ and $12$ basic VQC units (layers) and
the best outcomes are achieved for $m=8$. The latter
 are reported in Table \ref{tab:3}, where we also summarize the results of all models used in this study.

\begin{table}[t]
   
    \caption{ Metrics evaluating the  classification results of all models
    tested in this work. 
    For precision, recall and F1-score we list the macro average values.
    The number of total trainable parameters
    in each model is listed in the last column. For the QAE--Qudit model 490 from
    the total number of  parameters are attributed to the QAE model. For the QAE--Qubits model 220 parameters out of the 316 parameters  are used in the QAE model.}
     \centering
    \label{tab:3}
    \footnotesize 
    \begin{tabular}{|l||c|c|c|c|c|}
        \hline
      Model  &   Precision & Recall & F1 & Accuracy & $\#$ weights  \\
        \hline \hline
     QAE--Qudit & 0.93 & 0.82 & 0.84 & 0.93 & 1,130  \\
     \hline
     Classic NN & 0.94 & 0.83 & 0.85 & 0.94 & 67,328 \\
     \hline
     Qudit & 0.85 & 0.78 & 0.80 & 0.91 & 640 \\
     \hline
     QAE--Qubits & 0.35 & 0.42  & 0.38 & 0.62&  316 \\

        \hline
      \end{tabular}
\end{table}

\begin{figure}[!t]
    \centering
 \includegraphics[width=0.9\linewidth]{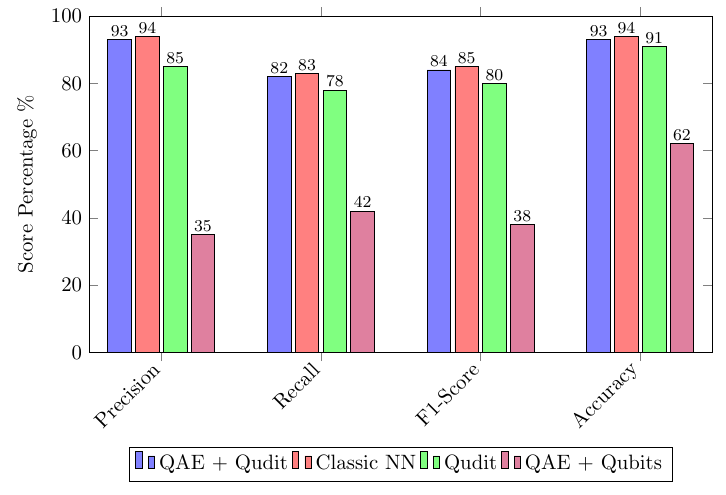}
    \caption{ A visual representation of the metric values for the classification task achieved by every  model under study in this work. The numerical valued for this plot are listed on Table~\ref{tab:3}.}
    \label{fig:3}
\end{figure}

\section{Discussion \label{S6}}

In this work, we developed a hybrid model capable of processing a large dataset consisting of around 10,000 rows, which is significantly larger than datasets typically explored in QML. The results achieved are comparable to those obtained with a deep NN, whose number of trainable weights is nearly two orders of magnitude greater than in the proposed (QAE--qudit VQC) model. These findings suggest that if the qudit-based VQC were a physical system, rather than being simulated as in this work, its training could potentially be faster than that of a deep NN

The comparison of the suggested model, the QAE-qudit VQC, with simpler models in Section~\ref{S6} suggests that there are two key elements that warrant consideration. The first is the use of an encoding adapted for the quantum circuit, via QAE. Incorporating the quantum feature map into its bottleneck, QAE appears to provide, in agreement with previous studies \cite{Maragkopoulos:2024QAE}, a boost in the accuracy of classification of approximately $3-4$~$\%$. The second key element is the absence of an ansatz in the VQC architecture. By employing a qudit, we are able to treat all generators of the qudit's symmetry group on an equal footing and construct the VQC in a uniform manner with respect to the allowed operations. Additionally, drawing on insights from previous research \cite{Mandilara2024}, we have used a compact form that integrates the variational part of the circuit and embedding of features into a single action. We believe that all these elements merit further theoretical investigation and additional testing on larger datasets to draw more robust conclusions.

\section*{Acknowledgements}
This work was supported by European Union’s Horizon
Europe research and innovation program under grant agreement No.101092766 (ALLEGRO Project) and  Hellas QCI project co-funded by the European Union under the Digital Europe Programme  grant agreement  No.101091504.

\bibliography{bibliography}

\end{document}